\documentclass[preprint,12pt]{elsarticle}

\usepackage{graphicx}
\usepackage{dcolumn}
\usepackage{bm}
\usepackage{amsmath}
\usepackage{amssymb}
\usepackage{amsfonts}
\usepackage{epsfig}
\begin{document}
\begin{frontmatter}

\title{Quantifying randomness in protein protein interaction networks of
different species: A random matrix approach}

\author[IITI]{Ankit Agrawal}
\author[IITI]{Camellia Sarkar}
\author[IITI]{Sanjiv K. Dwivedi}
\author[Okinawa]{Nitesh Dhasmana}
\author[IITI]{Sarika Jalan\corref{corr}}
\address[IITI]{Complex Systems Lab,
Indian Institute of Technology Indore, M-Block, IET-DAVV Campus Khandwa Road, Indore-452017}
\address[Okinawa]{Light Matter Interactions Unit, Okinawa Institute of Science and Technology, Okinawa, Japan}
\cortext[corr]{Corresponding author: sarika@iiti.ac.in}

\date{\today}

\begin{abstract}
We analyze protein-protein interaction networks for six different species under the framework 
of random matrix theory. Nearest neighbor spacing distribution of the eigenvalues of adjacency 
matrices of the largest connected part of these networks emulate universal Gaussian orthogonal 
statistics of random matrix theory. 
We demonstrate that spectral rigidity, which quantifies long range correlations in 
eigenvalues, for all protein-protein interaction networks follow random matrix prediction up 
to certain ranges indicating randomness in interactions. After this range, deviation from 
the universality evinces underlying structural features in network. 
\end{abstract}
\begin{keyword}
Random matrix theory, Protein Protein interaction networks, Spectra
\end{keyword}
\end{frontmatter}

\section{Introduction}
Random matrix theory (RMT), proposed by Wigner to explain the statistical 
properties of nuclear spectra, has elucidated a remarkable success in
understanding complex systems 
which include disordered systems, quantum chaotic systems, spectra of large complex atoms, etc. 
Recently complex networks have also been analyzed in RMT framework bringing them into
the universality class of Gaussian orthogonal ensemble (GOE) \cite{SJ_pre2007a,SJ_pre2007b,SJ_phy2008,SJ_geneexpression}. Systematic
investigations performed on model networks establish correlation between their structural properties and
spectral properties inspected by RMT.
This paper validates the access of a mathematical
tool, RMT, to study 
protein-protein interaction (PPI) networks of different species, as model systems, under RMT framework. 
By interaction we mean that a protein may 
change conformation of another protein leading to a change in its affinity for different 
groups or may lead to addition or removal of a group in the molecule. The interaction is 
highly specific i.e. it can discriminate among thousands of different molecules in its 
environment and selectively interact with one or two \cite{Biochemistry_book}.

A network representation of PPI in addition to providing a better 
understanding of protein function, serves us with a powerful model of various functional 
pathways elucidating mechanics at cellular level \cite{ClusteringMethods_2006,rev_bio_net,Satorras_2003}. 
Recently it has been realized that analysis of a network representation of such interactions, 
in comparison to pairwise analysis, provides a much better understanding of the processes 
occurring in biological systems \cite{ppi1,ppi2,RandomWalks_2005,Krapivsky_pre2005,Berg_BMC_2004}. These studies
indicate a strong correlation between the interaction networks and expression properties of 
the proteins having similar expression dynamics  i.e.  they tend to form 
clusters of either static or dynamic proteins \cite{Static_2007,GeneInter_2008}. 
Furthermore, analysis of PPI networks have contributed in various 
disease related biological studies, for instance study of human interaction data and Alzheimer's 
disease proteins has enriched our knowledge about its protein targets \cite{alzheimer_2006}. 
Some of the PPI network studies reveal that
pathogens tend to interact with hub proteins and proteins that are central to many paths
in the network \cite{pathogens_2008}.

Analysis performed here involves construction of 
networks in such a way that any pair of proteins can achieve only two states 
i.e. either they are connected or
not connected. We demonstrate that nearest neighbor spacing distribution (NNSD) of PPI networks of 
different species exhibit a similar statistical behavior of RMT,
bringing them all under the same universality class. Furthermore, long range correlations in spectra display a wide
range of behaviors. 

\section{RMT and Networks - What is the connection?}
The random matrix approach
regarded the Hamiltonian of a heavy nucleus (which is 
very complex due to the complexity of interactions between various nucleons) as behaving like a random matrix
chosen from Gaussian orthogonal ensemble (GOE) having a probability density $P$. Energy levels were approximated by the 
eigenvalues of this matrix and their statistics were studied \cite{mehta}. The functional form of $P$ defines
the type of ensemble. Later this theory
was successfully applied in the study of spectra of different complex
systems including disordered systems, quantum chaotic systems,
spectra of large complex atoms etc \cite{rev_rmt}. RMT is also shown to be of great use while
understanding the statistical structure of the empirical cross-correlation matrices appearing in the
study of multivariate time series. The classical complex systems where RMT has been
successfully applied are stock market \cite{financial}; 
brain \cite{Seba}; 
patterns of atmospheric variability 
\cite{Sant}, physiology and DNA-binding proteins \cite{rmt-DNA} etc. Our previous studies elucidate that different  model networks, namely scale-free, small
world, random networks and modular networks ensue universal GOE statistics of RMT.
		
A network is represented in the form of an adjacency matrix in which a corresponding
element $A_{i,j}$ is set to 1 if $i$ node is connected with $j$ node and 0 otherwise. The set of 
eigenvalues of an adjacency matrix is called its spectrum. Spectrum of a network is related with its various 
topological properties \cite{spectra}. Spectral density of adjacency matrix of a random network 
reflects a semicircular law \cite{semicircular}, which interestingly is one of the properties of random 
matrix chosen from a Gaussian ensemble \cite{mehta}. Even though the analysis of some real world 
networks and various model networks capturing real world properties manifest
somewhat different spectral densities \cite{semicircular,Jost1}, our 
previous studies reveal the universal GOE statistics of eigenvalue fluctuations of 
the above mentioned networks  
\cite{SJ_pre2007a,SJ_random,SJ_modular}.
This similarity in the NNSD of spectra
of a random matrix and that of different model networks furnishes more insight into the physical significance of spectra of networks. These eigenvalues can be treated as elements 
signifying different topological states of a network or it can be said that the spectra of a network 
gives us information about the property which is used to define the entries of adjacency matrix 
(connections) of network, as in case of nuclear spectra, different energy levels of nucleus are approximated
by eigenvalues of a random matrix (representing Hamiltonian i.e. energy of system). One more aspect,
which this connection between RMT and networks demonstrate, is the existence of
some amount of randomness in these networks. 
We take our previous studies a step further and validate the applicability of
RMT on PPI networks, demonstrating that they all come under the same universality class.

\section{Data sources and network construction}
To construct PPI networks of different species, first interaction data is 
downloaded from publicly available data source DIP (database of 
interacting proteins) \cite{DIP}. 
DIP manages a database for experimentally determined protein interactions in all organisms.
It integrates information from different sources to create a single set of protein protein interaction.
Data from this source is widely used for various data analyses and biological studies \cite{mpg}.
In the database, each protein and each interaction is represented by a unique id, and for  
each interaction, information of interactors are given. 
With this information, an interaction network is constructed for a species. 
Here nodes of the network are represented by the proteins and a 
connection in the network corresponds to the interaction occurring between the two protein represented by 
two nodes. 
Next we detect largest connected cluster
in these networks. Corresponding adjacency matrix has a entry $A_{i,j}= 1$ if protein $i$ 
interacts with protein $j$ and 0 otherwise. 
PPI networks are undirected and so the adjacency matrix is symmetric entailing all real
eigenvalues. In all 
the networks, a protein interaction with itself, if any, are ignored. 
Below we provide a 
brief description of different species studied here.

\section{Method of Analysis}

\subsection{NNSD}
In the following, we introduce spacing 
distribution of random matrices. Let eigenvalues of a network be denoted
by $\lambda_{i}, i = 1, . . . , N$ 
and $\lambda_{1} < \lambda_{2} < \lambda_{3}<$ $\dots< \lambda_{N}$. 
In order to get universal properties of the fluctuations 
of eigenvalues, it is customary in RMT to unfold the eigenvalues by a transformation 
$\overline{\lambda_{i}} = \overline{{N}}(\lambda_{i})$, 
where $\overline{N}$ is average integrated eigenvalue density \cite{mehta}. Since we do not have any 
analytical form for $\overline{N}$, we numerically unfold the spectrum by polynomial curve fitting 
(for elaborate discussion on unfolding, see Ref.~\cite{mehta}). After unfolding, average spacing 
becomes unity, independent of the system. Using the unfolded
spectra, we calculate spacings as $s_1(i) = \overline{\lambda_{i+1}}-\overline{\lambda_{i}}$.
In the case of GOE statistics, the nearest neighbor spacing distribution (NNSD) is denoted by  
\begin{equation}
\label{first} 
P(s_1) = \frac{\pi}{2}s_1\exp\left(-\frac{{\pi}s_1^2}{4}\right).
\end{equation}

For intermediate cases, the spacing distribution is described by Brody parameter \cite{Brody}.
\begin{subequations}
\begin{align}
P_{\beta}(s_1)&=As_1^\beta\exp\left(-\alpha s_1^{\beta+1}\right)
\end{align}
where $A$ and $\alpha$ are determined by the parameter $\beta$ as follows:
\begin{align}
A&=(1+\beta)\alpha, \, \, \, \alpha=\left[{\Gamma{\left(\frac{\beta+2}{\beta+1} \right)  }}\right]^{\beta+1}
\end{align}
\label{eq_brody}
\end{subequations}
This is a semi-empirical formula characterized by parameter $\beta$. As $\beta$ goes from 0 to 1, the
Brody distribution smoothly changes from Poisson to GOE. We fit spacing distributions of different
networks by the Brody distribution $P_{\beta}(s)$. This fitting gives an estimation of $\beta$, and
consequently identifies whether the spacing distribution of a given network is Poisson, GOE, or the
intermediate of these two \cite{Brody}.

Apart from NNSD, the next nearest-neighbor spacing
distribution (nNNSD) is also used to characterize the
statistics of eigenvalue fluctuations. We calculate this
distribution $P(s_2)$ of next nearest-neighbor spacing,
\begin{equation}
s_2(i)
= (\lambda_{i+2} - \lambda_{i})/2
\label{eq_s2}
\end{equation}
between the unfolded eigenvalues. Factor of two at the
denominator is inserted to make the average of next nearest-
neighbor spacing $s_2(i)$ unity. According to Ref.
\cite{mehta}, the nNNSD of GOE matrices is identical to the NNSD
of Gaussian symplectic ensemble (GSE) matrices, i.e.,
\begin{equation}
P(s_2) = \frac{2^{18}}{3^6\pi^3}s^4_2 exp(-\frac{64}{9\pi} s_2^2)
\label{eq_nNNSD}
\end{equation}
The NNSD and nNNSD reflect only local correlations
among the eigenvalues. The spectral rigidity, measured
by the $\Delta_3$-statistics of RMT, preserves information about the
long-range correlations among eigenvalues and is a more
sensitive test for studying RMT properties of the matrix under
investigation. In the following, we describe the
procedure to calculate this quantity.

\subsection{$\Delta_3$ statistics}
The $\Delta_3$-statistics measures the least-square deviation of the spectral staircase function representing
average integrated eigenvalue density $\overline{N}(\lambda)$ from the best fitted straight line for
a finite interval of length $L$ of the spectrum given by
\begin{equation}
\Delta_3 (L;x) = \frac{1}{L} min_{a,b}   \int_x^{x+L} [N(\overline{\lambda})-a\overline{\lambda}-b]^2 d\overline{\lambda}
\label{eq_delta3}
\end{equation}
where $a$ and $b$ are regression coefficients obtained after least square fit. Average over several choices
 of x gives the spectral rigidity $\Delta_3(L)$. For GOE case, $\Delta_3(L)$ depends logarithmically on L, i.e.
\begin{equation}
\Delta_3(L)  \sim \frac{1}{\pi^2} \ln L. 
\label{eq_delta3_goe}
\end{equation}

\section{Results}
In this section we present various results obtained for each of the different PPI networks constructed.
All the results are produced for the adjacency matrix corresponding to the largest connected cluster of each 
network. For completeness we briefly discuss degree distribution and density distribution 
for all these species, after which present random matrix results.
\begin{figure}
\centering
\includegraphics[width=0.9\columnwidth]{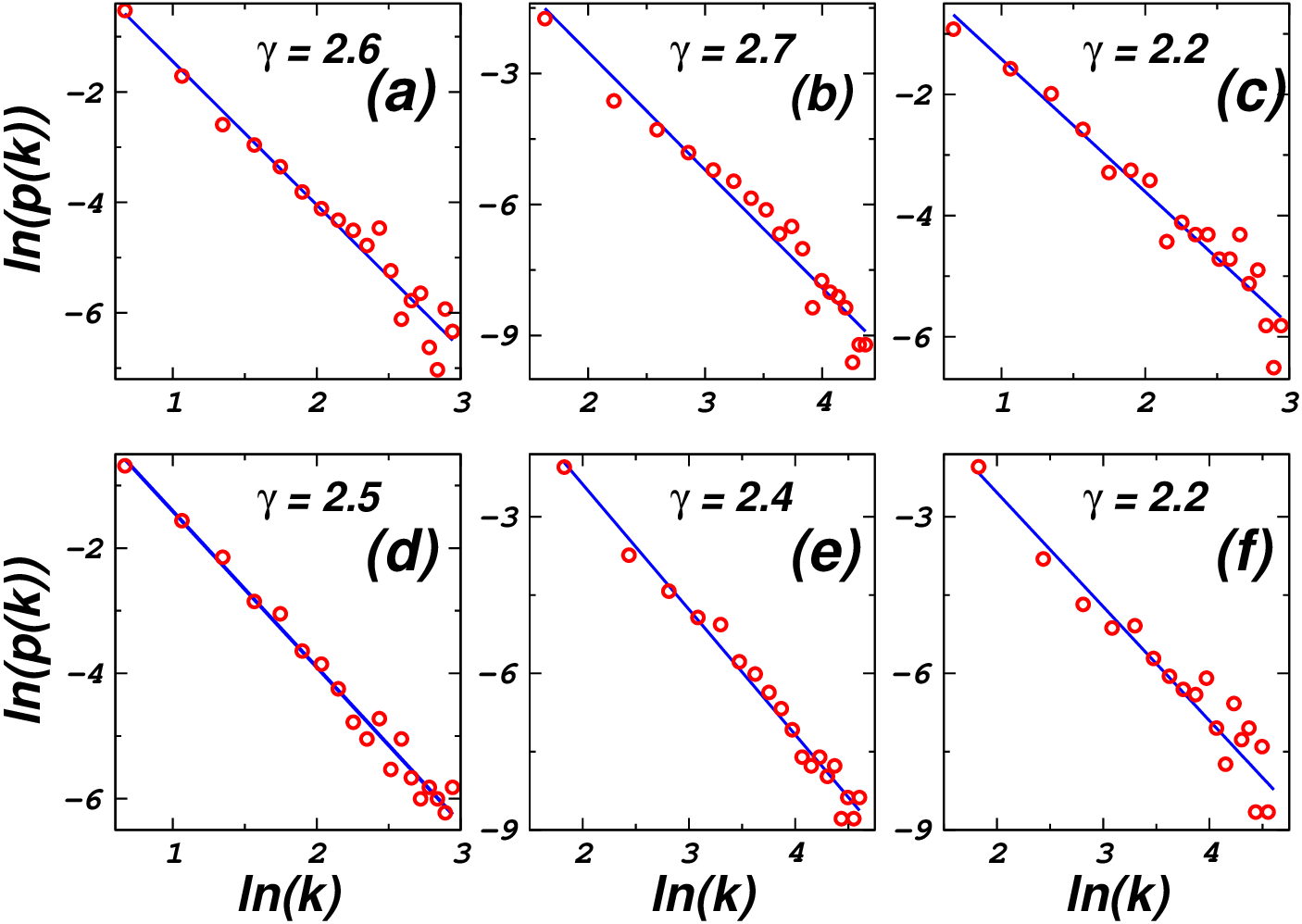}
\caption{(Color online) Degree distribution $p(k)$ of the largest connected network for all the species. 
(a) {\it C. elegans}, (b){\it D. melanogaster}, (c) {\it H. pylori}, (d) {\it H. sapiens}, (e) {\it S. cerevisiae} and (f) {\it E. coli}. 
The solid line (blue) obtained from the fitting to the power law function
$p(k) \sim k^{-\gamma}$ where $\gamma$ is a constant.
$\gamma$ value for all plots come out to be between 2 and 3, indicating scale-free nature.}
\end{figure}

\subsection{Degree Distribution}
Fig.~(1) plots degree distribution $\rho(k)$
for each of the PPI networks for the six
species studied and corresponding $\gamma$ values are obtained. The results confirm that all the 
PPI networks are scale-free entailing power law \cite{rev_network}. 

\subsection{Spectral Analysis}
\begin{figure}[h]
\centering
\includegraphics[width=0.9\columnwidth]{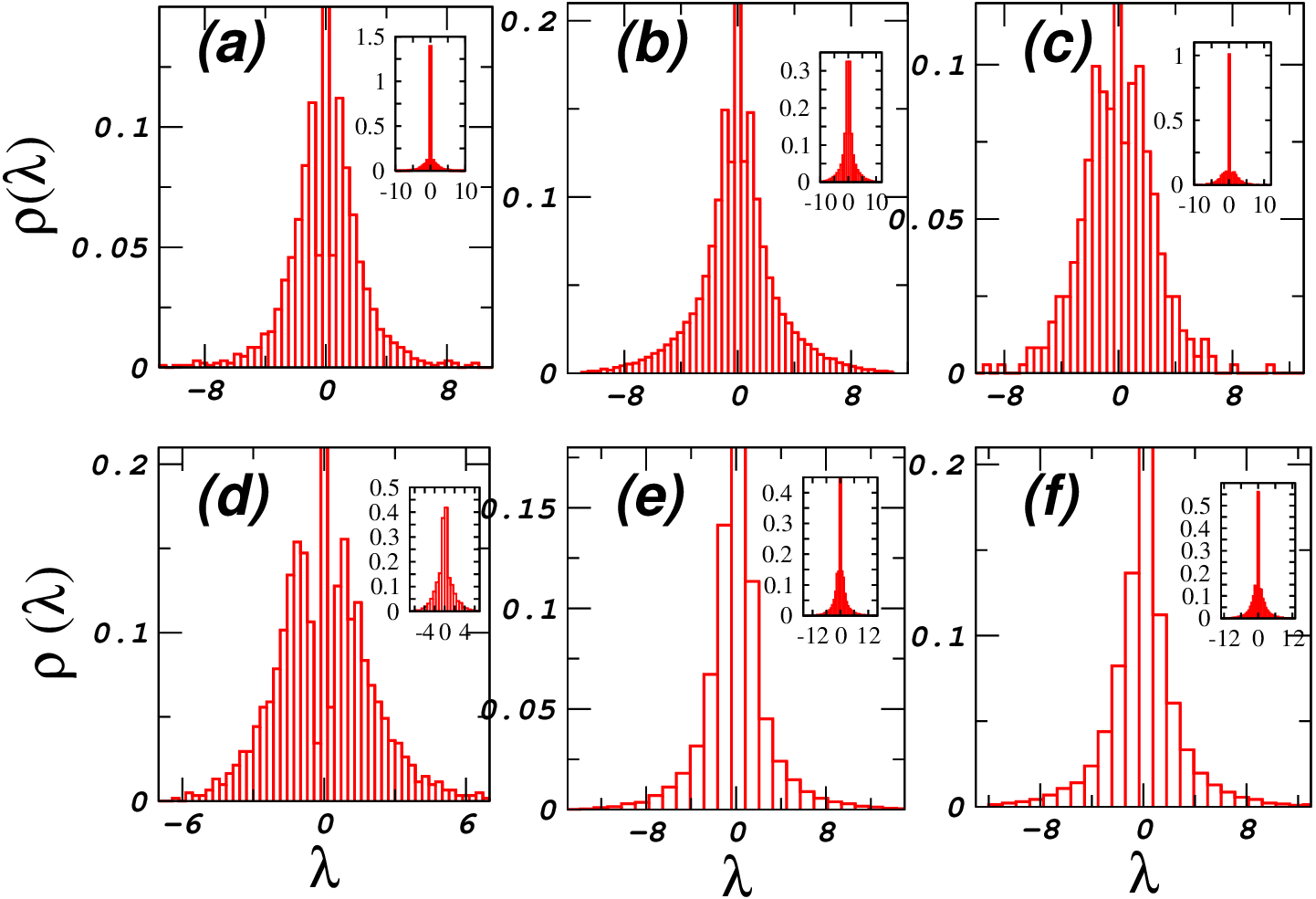}
\caption{(Color online) Eigenvalue distribution $\rho(\lambda$) for largest 
connected networks of all species [(a)-(f) for {\it C. elegans}, {\it D. melanogaster},
 {\it H. pylori}, {\it H. sapiens}, {\it S. cerevisiae}, {\it E. coli} respectively]. Inset depicting peak of distribution.}
\label{fig_eigen}
\end{figure}

The PPI networks considered here
are undirected entailing all real eigenvalues. The spacing between the adjacent smallest eigenvalues
is very large which sharply decreases initially, then gradually falls to zero followed by a
sharp increase. The plots do not provide a quantitative estimate that how similar 
eigenvalues are spaced together in different networks but more importantly it convey the 
information
about degeneracy in the network which has to been taken into account for
NNSD analysis. Real world networks, in general, are very sparse and are
reported to have a large number of zero eigenvalues\cite{sparse_2005}. 
Fig.~\ref{fig_eigen} plots 
normalized spectral density $\rho(\lambda)$ for each of the networks. Profile for the 
spectral density 
looks more like bell shaped for all the networks. Even though the global nature of variation of spectral 
density with the eigenvalues looks similar in different networks, minute observations of 
the plot in the inset of spectral density figures indicate that they vary differently in different networks. 
\subsection{Short range correlations in eigenvalues}
\begin{figure}
\centering
\includegraphics[width=0.8\columnwidth]{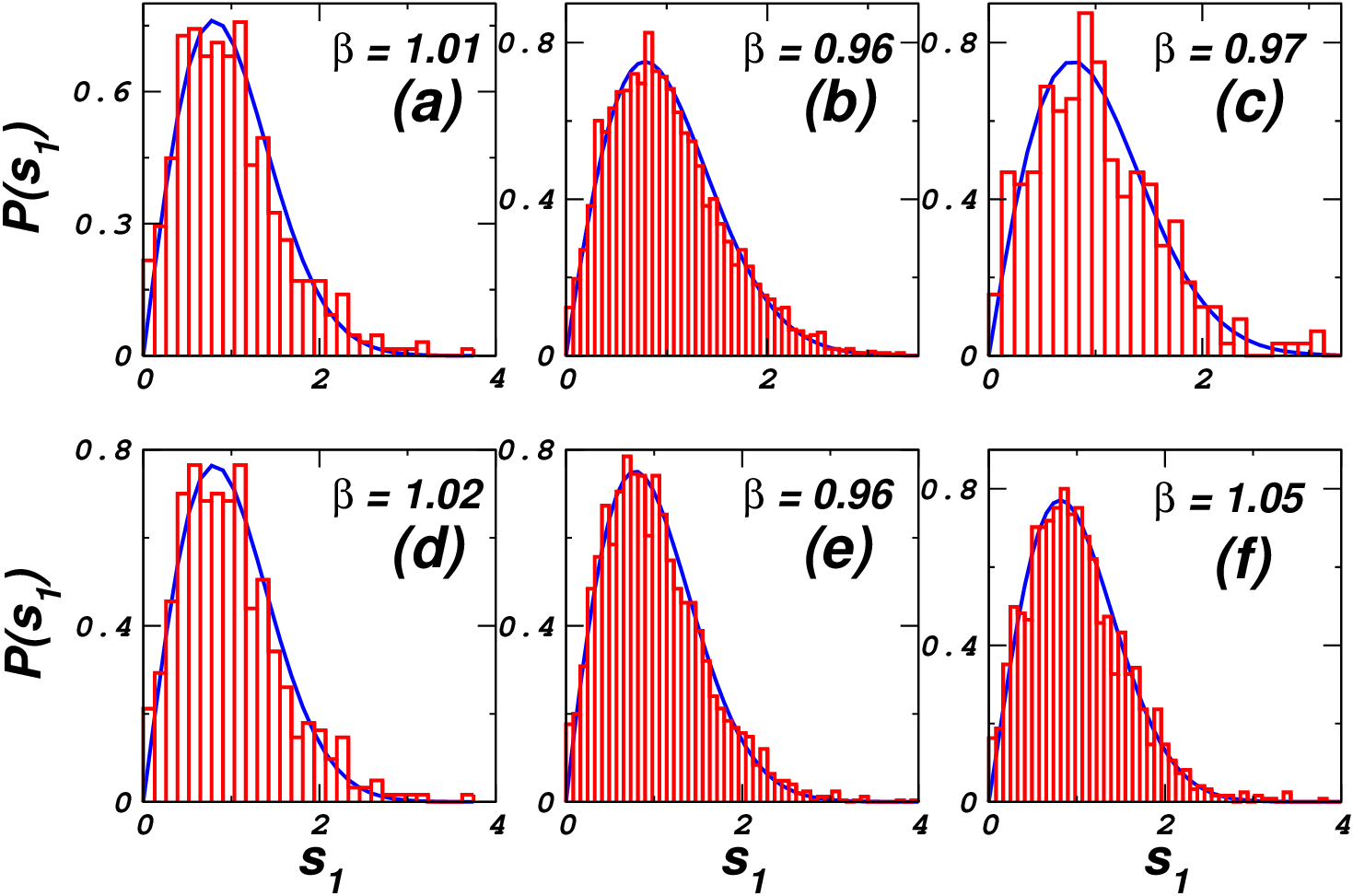}
\caption{(Color online) Nearest neighbor spacing distribution (NNSD) $P(s_1)$ of the adjacency 
matrices of different species of networks. (a) {\it C. elegans}, (b) {\it D. melanogaster}, 
(c) {\it H. pylori}, (d) {\it H. sapiens}, (e) {\it S. cerevisiae} and (f) {\it E. coli}. All emulate
GOE statistics. 
The histograms are numerical results and the solid lines represent fitted Brody distribution
 (Eq.~\ref{eq_brody}). The value of $\beta$ close to $1$ corresponds to GOE distribution.}
\label{fig_nnsd}
\end{figure}

From the spectra of each of the networks we calculate the spacing distribution for adjacent
eigenvalues. Negative and positive eigenvalues are unfolded separately with different
polynomial functions. The degenerate eigenvalues, other than zero, are
considered as a single eigenvalue. The flat region corresponding to zero eigenvalue is 
excluded from the analysis. Extreme eigenvalues at both the ends of the spectra
are also not considered. Fig.~\ref{fig_nnsd} plots NNSD for different networks 
corresponding to different species. All the plots are fitted with the Brody distribution
given by Eq.\ref{eq_brody}. 

The value of fitted Brody parameter indicates that the NNSD of eigenvalues for 
all the PPI networks ensue GOE distribution. 
This is not a trivial result and it renders a new look into such interaction
networks, as in spite of genetic differences, differences in internal environment, 
biological activities, modes of functioning in those species affecting their protein-protein interactions,
they exhibit a similar universal behavior predicted by RMT.
Another RMT interpretation of GOE statistics suggests that some amount of randomness
is present in PPI networks. 
\begin{figure}
\includegraphics[width=\columnwidth]{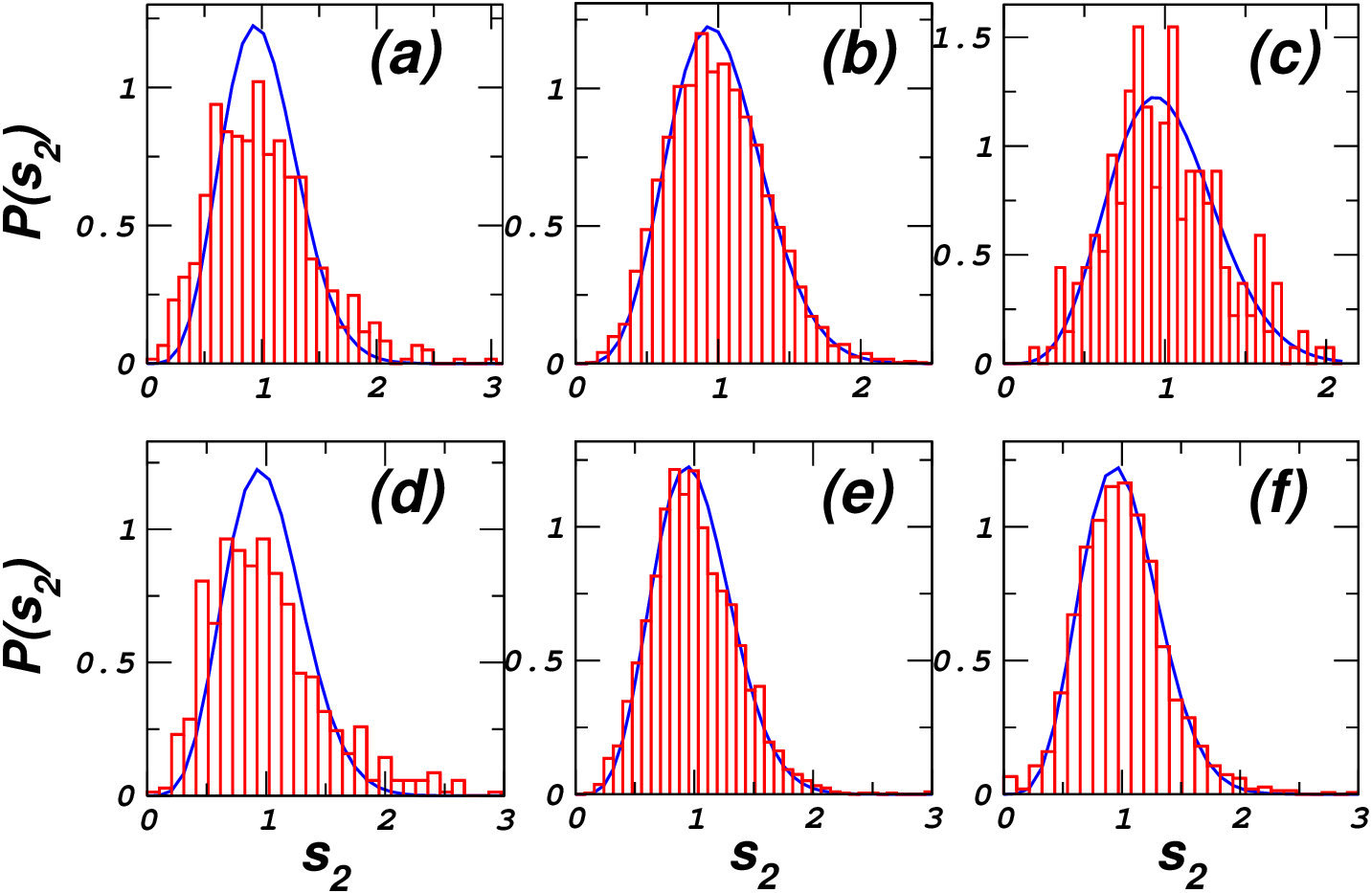}
\caption{(Color online) Next nearest-neighbor spacing distribution (nNNSD) $P(s_2)$ of the 
adjacency matrices of largest connected network of each species.
(a) {\it C. elegans}, (b) {\it D. melanogaster}, (c) {\it H. pylori}, (d) {\it H. sapiens}, (e) {\it S. cerevisiae} 
and (f) {\it E. coli} are compared with the nearest neighbor spacing distribution (NNSD) of GSE matrices.
Histograms are numerical results and solid lines represent NNSD of GSE matrices 
(Eq.\ref{eq_nNNSD}).} 
\label{fig_s2}
\end{figure}
Same unfolded eigenvalues have been used to calculate next nearest-neighbor
spacings as in Eq.~\ref{eq_s2}. nNNSD for all PPI networks are
plotted in Fig.~\ref{fig_s2}.
For {\it C. elegans} and {\it H. sapiens}, though NNSD emulates GOE statistics of RMT, nNNSD shows deviation
from it, this supports that NNSD is not very sensitive test of RMT, and in order
to learn more about correlations in eigenvalues one should go for other tests like
nNNSD or $\Delta_3$ statistics. For other networks, next nearest-neighbor spacings
confirm random matrix predictions of GOE statistics.
\subsection{Long range correlations in eigenvalues}
\begin{table}[!t]
\begin{center}
\caption{For each species, $N_{-}$, $N_{+}$ and $N_{0}$ represent number of eigenvalues corresponding to negative, positive and zero regions, respectively.  
$N_{ori}$ is size of the original network downloaded from the dataset,
while $N$ is largest connected component for each species. $L$ is the length of spectrum upto
which statistics comply with RMT. Values of $L$ have not been included for species which do not follow 
$\Delta_3(L)$ statistics.
}
\begin{tabular}{|l|l|l|l|l|l|l|l|}	\hline
Species & $N_{ori}$ & N & $N_{+}$  & $N_{-}$  & $N_{0}$  & L & \% L/N \\ \hline
{\it C. elegans}  & 2646 & 2386 & 516 & 516 & 1354 & - & -	\\ \hline
{\it D. melanogaster} & 7451 & 7321 & 2504 & 2506 & 2311 &	28 & 0.38	\\ \hline
{\it H. pylori}    &  732  & 709 & 196 & 196 & 317  &	13 &  1.83	\\ \hline
{\it H. sapiens}  & 3164  & 2138 & 628 & 646 & 864 & - & -		\\ \hline
{\it S. cerevisiae} & 5080 & 5019 & 1995 & 2048 & 976 & 18 & 0.35	\\ \hline
{\it E. coli} & 2969 & 2209  &  851 & 871 & 487 & 9 & 0.41 \\ \hline
\end{tabular}
\end{center}
\label{table}
\end{table}
\begin{figure}
\centering
\includegraphics[width=\columnwidth]{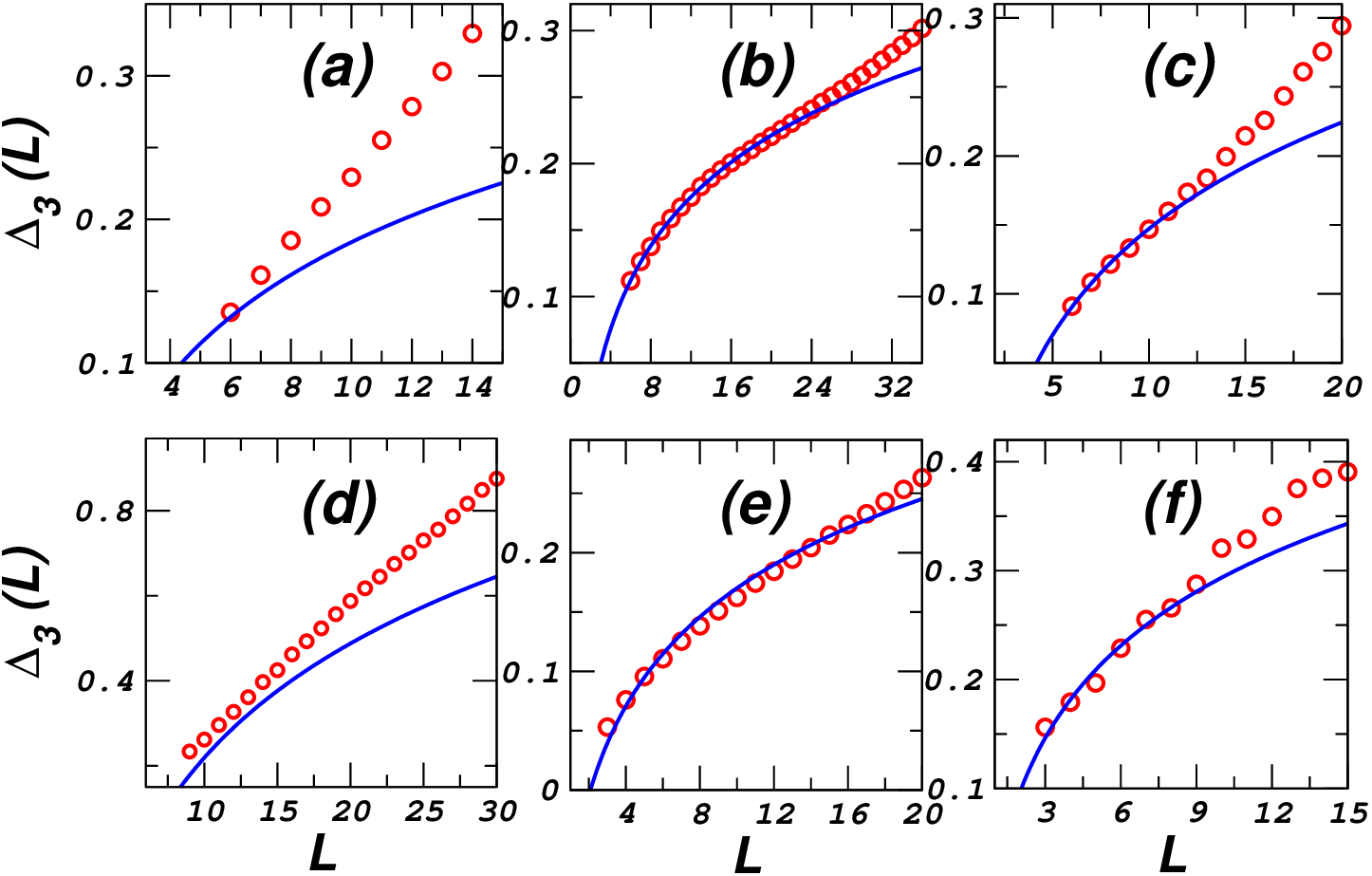}
\caption{(Color online) $\Delta_3$ (L) statistics for largest connected 
networks of each species (open circles).
(a) {\it C. elegans}, (b) {\it D. melanogaster}, (c) {\it H. pylori}, (d) {\it H. sapiens}, 
(e) {\it S. cerevisiae} and (f) {\it E. coli}. The solid line represents the GOE prediction.  
$\Delta_3(L)$ statistics follows the RMT prediction up to length $L$ which is 
$\sim$ 6, 28, 13, 9, 18 and 9 for the respective figure [(a)-(f)].}
\label{fig_delta3}
\end{figure}
As explained in the introduction that NNSD and nNNSD offer insight into only short range correlations among the eigenvalues. 
To probe for long range correlations one employs $\Delta_3$ statistics of the
spectrum of a network. For all the networks, $\Delta_3$ statistics is calculated
using Eq.~(\ref{eq_delta3}). Fig.~(\ref{fig_delta3}) elucidates this statistics
for various PPI networks, indicating that RMT does not provide a good model
for PPI networks of {\it C. elegans} and {\it H. sapiens}. Whereas, other PPI networks conform with $\Delta_3$ statistics
of GOE up to a certain range $L$, after which deviation from this universal GOE statistics
occurs. According to RMT,
eigenvalues are correlated up to this range.
Different ranges of $L$ in different species,
for which $\Delta_3$ statistics imitates RMT, can be understood as different amount of
randomness in corresponding PPI network \cite{SJ_random}. Last column of the Table.~1 
displays a qualitative measure of randomness in network as detected in eigenvalue
correlations.

\section{Summary of results for different species}
{\it C. elegans} - The eigenvalues vary in the interval $(-14,16)$. Degeneracy in three 
regions is observed for $\lambda$=-1, 0 and 1. Out of the 2386 eigenvalues, nearly 56\% of the 
eigenvalues are zero. Although the degeneracy at -1 and 1 is considerably less as compared to zero,
$0.7\%$ of the eigenvalues are degenerate for each of $\lambda$=-1 and $\lambda$=1. The degree
 distribution for the corresponding network clearly reveals that it is scale-free with
$\gamma$ = 2.6.  The normalized spectral density, on the other hand, shows a little deviation
 from the triangular distribution. Spectral density as expected is very large around 0 and decreases as
 the magnitude of eigenvalue increases. The universality of {\it C. elegans} PPI networks is quantified by the NNSD
 analysis. The quantifying factor is the Brody parameter which for this network comes out to be
 $\beta=1$, but nNNSD and $\Delta_3$ indicate deviations. There can be two interpretations of this 
behavior. First one is that PPI
network of {\it C. elegans} cannot be modeled using GOE of RMT, as NNSD complying GOE statistics is not
very sensitive or reliable test. Second interpretation is that there is a very minimal amount of 
randomness in underlying matrix bringing upon correlations between only nearest
neighbors in spectra.

{\it D. melanogaster} - The eigenvalues vary in the interval $(-15,22)$. Out of the 7321 eigenvalues, 31\% are 0, only 9 and 5 eigenvalues show degeneracy for $\lambda$=-1 and 1, respectively. Thus,
 the amount of degeneracy in this species is considerably less as compared to {\it C. elegans} and the plot
 for eigenvalues is smoother as well. This network also fits well with the power law degree distribution
 ($\gamma$ = 2.7), verifying the scale-free nature for this network as well. Normalized spectral density
 for this network also resembles the bell-shaped curve but the density as compared to that of {\it C. elegans} decreases
 much slower as the magnitude of eigenvalues increases. The NNSD fitting with the Brody Distribution gives
 $\beta \sim 0.96$, which clearly indicates the GOE behavior of NNSD.
nNNSD too abides well with GOE predictions of RMT. Long range correlations measured by $\Delta_3$ statistics
agrees with RMT prediction up to $L \sim 28$.

{\it H. pylori} - The 709 eigenvalues vary between $(-9,10)$. Degeneracy, in this case, is observed only
 for $\lambda$=0, with $44\%$ of the eigenvalues being 0. Degree distribution found to obey 
power law verifying its class of scale-free networks. Corresponding $\gamma$ value is 
2.2. The normalized spectral density for network of this species is more triangular than the above two species. 
Brody fitting to NNSD yields $\beta=0.97$. nNNSD agrees well with the NNSD of GSE matrices. 
Long range correlation ($\Delta_3$ statistics) ensues GOE statistics up to length $L \sim 13$. 

{\it H. sapiens} - Network for {\it H. sapiens} exhibits three types of degeneracy. 40\% of the 2138 eigenvalues
are 0 while 1.5\% of eigenvalues corresponds to $\lambda$=-1 and only 0.7\% values show degeneracy at
$\lambda$=1. Eigenvalues vary in the interval $(-9,11)$. Degree distribution fits well with the power
law giving value of $\gamma$ = 2.5. Normalized spectral density for this species is also  
bell shaped but slopes of the two curves are sharper than the previous networks bringing more
triangularity in them. Also the density is irregular around zero. This irregularity in the density may be
because of the missing interaction knowledge in human beings. Again NNSD
fitted well with Brody statistics with $\beta\sim 1$, indicates GOE statistics, but nNNSD
does not conform with GOE predictions. 

{\it S. cerevisiae} (baker's yeast) - Eigenvalue degeneracy is observed at $-1$, $0$ and $1$. Of the 5018 eigenvalues,
 20\% are 0, while degeneracy at $-1$ and $1$ makes up around 0.43\% and 0.23\%, respectively. The eigenvalues in this network vary in the interval $(-32,40)$. The zero degeneracy is minimum in this network which
may shed some light into its evolution \cite{Jost1}. High protein-protein interaction knowledge persists
about yeast as it is the most extensively studied species.
Value of $\gamma$ obtained by
 fitting the degree distribution with the power law is 2.4. Fitted Brody parameter for network of 
this species comes out as 0.96 which brings this species as well under the universality class 
emulating GOE statistics \cite{Phys_Lett2006}. nNNSD too agrees well with the NNSD of GSE matrices. Long range
correlations measured by $\Delta_3$ statistics agrees well with the RMT prediction up to
length $L \sim 18$. 

{\it E. coli} - Eigenvalues lie between $(-30,46)$, out of which $0.4\%$ eigenvalues are having 
degeneracy at $-1$, $0.1\%$ at 1 while 22\% of eigenvalues makes up the degeneracy at 0. The total number of 
eigenvalues is 2209. Zero
 degeneracy in this species is also considerably less. Degree distribution for the network assimilates 
power law verifying its scale-free nature. Value of $\gamma$ obtained is 2.2. NNSD reflects GOE statistics 
with Brody parameter equal to $\beta \sim 1$. nNNSD too agrees
with the GOE predictions. Long range correlations measured by $\Delta_3$ statistics agrees 
substantially with RMT prediction up to length $L \sim 9$.  

\section{Conclusions and discussion}
Taking this journey of understanding how nature works further, considering the fact that almost
 all the biological processes in all the organisms occur via different protein-protein interactions,
 we demonstrate universality of such interactions in the species studied in this paper. The Brody 
parameter for the nearest neighbor spacing distribution has value near $one$ which corresponds to  
the GOE distribution. We attribute this universality to the existence of {\it minimal amount of 
randomness} in all these networks.
The diversified nature of set of species studied makes it highly probable that this universality 
is also found in other organisms for which interaction data is not yet available. With the 
increase in interaction data of different species, by using the analysis carried out in this paper, it can be verified 
whether the universality is global or not. 

Earlier studies have reported that NNSD of biological networks
exhibit GOE statistics indicating short range correlations
in eigenvalues, hence demonstrating the applicability of RMT.
We carry investigation of biological networks under RMT framework further
by analyzing long range correlations in eigenvalues of PPI networks, and
report that $\Delta_3$ statistics of different networks emulate GOE statistics of RMT
for different ranges. Different PPI spectra elucidating short range correlations in spectra
displayed by GOE statistics of NNSD is somewhat obvious as underlying network is complex and random.
But deviation from universality at next to next nearest spacing itself, as displayed by two
species indicating deviation from randomness, stems many intriguing questions. For instance, how, in the course of evolution,
the interaction network discerns that it has sufficient randomness to introduce short
range correlations in corresponding spectra leading to apparent suppression in random mutation?
Since, in evolutionary science, mutations are known to create heritable variations that are abundant, random and undirected. 
Natural selection directs evolution by sorting the initially random mutation variants according to
their adaptive values. These favorable variants might apparently be responsible for spreading
of sufficient randomness in terms of GOE statistics entailed by the spectrum, evident in the biological systems under study.
This assimilation is seeded from our earlier investigations reporting that spectra of model networks exhibit transition to GOE statistics as random connections are introduced
in underlying network \cite{SJ_pre2007a}, incorporating a transition to the small-world phenomena defined by small diameter and large clustering coefficient \cite{Strogatz_Nature1998}. Furthermore, this model network reflects GOE statistics upto length $L$ having direct correlation with number of random rewirings
\cite{SJ_random}. These findings based on model systems can be comprehended for biological networks investigated here as following; biological systems are just sufficiently random enough to confer robustness to their systems.
Moreover, mutations in biological systems can be considered as a means of introducing just sufficient randomness which is required to
introduce small range correlations in spectra yielding universal GOE statistics. 
After this minimal amount of randomness, which could be interpreted as resulting from random mutations, further mutation 
may be non-random as put forward by many researchers in evolutionary biology \cite{DNA_loss_Drosophila2002}.
Based on RMT, we construe this in terms of the deviation observed from the universality captured through 
spectral rigidity and nNNSD  statistics.
Two of the PPI networks investigated here exhibit deviation from universal GOE predictions even for nNNSD statistics, 
which provides a very good example of model systems from random matrix point of view. From the point of view of
biological evolution, it can be considered as a supportive evidence of non-random mutations prevalent in biological systems.  

We would like to make a note here that randomness here in networks are altogether different from the concept of
noise in dynamical systems. Randomness in  networks is referred to as random connections between nodes, 
which probably arise in the course of evolution randomly and may not be because of a particular 
functional role or importance of that connection.  For example, the module structure in networks 
is known to have specific functional motivation in the evolution and various modules could be 
considered to be linked with each other through random connections \cite{Pereira_2006}. 
Furthermore, protein-protein interaction databases have been reported in literature \cite{von_2002} to be incomplete
containing false positive links.
Several attempts have been made to determine
the impact of incompleteness and noise in data on the results of the analysis conducted on such
datasets. There have been enormous discussions on the degree distribution aspect of real world
networks generated using experimental or empirical data \cite{Stumpf_2005a}. 
These studies suggest that
false positives of PPI data appear to affect
network alignments little compared to false negatives indicating that incompleteness,
not spurious links \cite{barabasi_2003}, is the major challenge for interactome-level comparisons \cite{PPI_book}.
The present paper focuses on spectral properties of protein-protein interaction networks,
and results of NNSD are robust to false negative as well false positive links. However, the long range correlations
would exhibit
dependence on false positive links as these links can be considered as `random connections' in networks
and more false positive links would lead to a larger range of $L$ for which $\Delta_3(L)$ statistics would
follow random matrix theory. The implication of this fact is that we cannot compare randomness of two
networks having very closeby values of $L$ but false positive links would not pose any problem for networks differing in $L$ significantly. In order to investigate the effect of false positive links on spacing distribution,
we introduced 5\% false connections randomly to the networks generated from the data downloaded
from DIP for the six species under investigation. The NNSD, thus generated does not show noticeable
variation from that generated from original datasets exemplifying the fact that the system is robust
against deleterious perturbations arising from false positive links. Similar results were observed on
random removal of 5\% links from the original datasets, where NNSD of the hence generated networks for
all species still follow GOE statistics.

Though we are far from making a strong conclusion based on range for which spectrum follows RMT,
as much of the random matrix results presented here cannot be interpreted along the lines of random 
matrix interpretation of dynamical systems, the universality observed in all networks prepares a platform to study these networks 
under RMT, focusing on randomness present in underlying systems. 
Randomness in biological systems is an essential component of heterogeneous determination and also 
acts as a key component of its structural stability owing to interactions between various levels of 
organization \cite{randomness_2013}. This extra dimension integrated with the results of network theory and already 
known biological knowledge can lead to extraction of useful information out of these complicated systems. 
We believe that nature adopts some algorithm to perform its functions. RMT uses properties of
random matrices to explain interactions in complex systems, systems which are although complex 
to study but are deterministic and governed by physical laws, demonstrating a non-random behavior in the
concerned systems after a certain extent. The applicability of RMT in protein interaction networks 
and deviation from universality imparts 
an interpretation of deterministic nature of protein interaction networks together with its 
complexity \cite{SJ_random}. This interesting phenomenon in biological systems pertaining to a sustained 
degree of randomness can be probed further to construct artificial systems reflecting such universality in their behavior. 

\section*{Acknowledgements} SJ thanks DST and CSIR for funding. SKD acknowledges UGC for financial support.

\section{Appendix}
The clustering coefficients and diameter for PPT networks are compared with those of the corresponding random 
networks. $C_{random}$ and $D_{random}$ have been calculated as {\it $\langle k\rangle$/n} and 
ln({\it n})/ln({\it $\langle k\rangle$}) \cite{Strogatz_Nature1998}. 
The PPI networks investigated in this paper are observed to have much higher values
of clustering coefficient than their 
corresponding random networks, whereas diameter of PPI networks are very close to those of random
ones, indicating their small-world nature \cite{Strogatz_Nature1998}.

\begin{table*}[!t]
\begin{center}
\caption{ $N$ is largest connected component for each species. $\langle k\rangle$, C, D denote average degree, clustering coefficient and diameter of the PPI (model) networks, respectively.$C (C_{random})$ and $D (D_{random})$ are the clustering coefficient and 
diameter of PPI (random) networks, respectively.}
\begin{tabular}{|l|l| l|l|l|l|l |l|}	\hline
Species 	& N &  $\langle k\rangle$ 	&  C   &      D &  $C_{random}$   & $D_{random}$ \\ \hline
{\it C. elegans}  	&  2386 & 3.206  	 	& 0.022 &  14 &  0.00134	&  7	\\ \hline
{\it D. melanogaster} &  7321 & 6.159   		& 0.011	&  11	&  0.00084	&  5		\\ \hline
{\it H. pylori}    	&  709  & 3.935   		& 0.015	&  9  &  0.00555	&  5		\\ \hline
{\it H. sapiens}  	&  2138 & 2.872  	 	& 0.111	&  25 &  0.00134	&  7		\\ \hline
{\it S. cerevisiae} 	&  5019 & 8.803   		& 0.097	&  10 &  0.00175	&  4   		\\ \hline
{\it E. coli}		&  2209 & 9.895   		& 0.109	&  12	&  0.00448	&  3	        \\ \hline
\end{tabular}
\end{center}
\label{table2}
\end{table*}

\end{document}